# Direct evidence of electronic interaction at the atomic-layer-deposited MoS$_2$ monolayer/SiO$_2$ interface

*Minji Lee[⊥], Yejin Kim[†], Ahmed Yousef Mohamed[†], Han-Koo Lee[∥], Kyuwook Ihm[∥], Dae Hyun Kim[§], Tae Joo Park[⊥,§], Deok-Yong Cho[†,\*]*

[⊥]Department of Materials Science & Chemical Engineering, Hanyang University, Ansan 15588, Korea, [†]IPIT and Department of Physics, Jeonbuk National University, Jeonju 54896, Korea, [∥]Pohang Accelerator Laboratory, Pohang 37673, South Korea, [§]Department of Advanced Materials Engineering, Hanyang University, Ansan 15588, Korea

KEYWORDS. MoS$_2$; SiO$_2$; interface; van der Waals interaction; electronic structure; X-ray absorption spectroscopy; synchrotron X-ray photoelectron spectroscopy

ABSTRACT

The electronic structure of an atomic-layer-deposited MoS$_2$ monolayer on SiO$_2$ was investigated using X-ray absorption spectroscopy (XAS) and synchrotron X-ray photoelectron spectroscopy (XPS). The angle-dependent evolution of the XAS spectra and the photon-energy-dependent evolution of the XPS spectra were analyzed in detail using an ab-initio electronic structure simulation. Although similar to the theoretical spectra of an ideal free-standing MoS$_2$ ML, the experimental spectra exhibit features that are distinct from those of an ideal ML, which can be interpreted as a consequence of S–O van der Waals (vdW) interactions. The strong consensus among the experimental and theoretical spectra suggests that the vdW interactions between MoS$_2$ and adjacent SiO$_2$ layers can influence the electronic structure of the system, manifesting a substantial electronic interaction at the MoS$_2$–SiO$_2$ interface.

**Introduction**

MoS$_2$, a representative two-dimensional (2D) transition metal dichalcogenide (TMD), is attracting significant attention regarding its excellent hydrogen evolution reaction (HER) capability [1-9]. Its catalytic HER activity is known to be dominated by electron exchange at the surface or edges of the specimen [3-5,10,11], implying the significance of the surface/interface electrical properties in determining chemical functionality. Recent advances in the fabrication of large MoS$_2$ monolayer (ML) flakes and in the artificial phase control from semiconducting 2H phase to metallic 1T phase [12-16] may expedite the commercialization of the TMD-based catalysis.

Semiconducting or metallic 2D TMDs are usually deposited on an insulating buffer, most likely SiO$_2$ [17,18], with a very large bandgap (> 9eV) to minimize the parasitic leakage through the bottom oxide and secure the functionality of the 2D electronic structure. The mechanism of adhesion of the MoS$_2$ ML onto SiO$_2$ has been regarded simply as a weak van der Waals (vdW) interaction, so no substantial chemical interactions are expected [19,20]. However, a clear examination of the significance of the vdW interaction on the electronic structure of the MoS$_2$ ML has not yet been performed. One of the major reasons is the difficulty involved in clear experimental verification of the interfacial influence due to the large deviation in the electronic properties of MoS$_2$ itself when multilayer MoS$_2$ is partially formed among the flakes.

Recent successes in ML fabrication using atomic layer deposition (ALD) have shed light on the precise quantification of the interfacial interactions owing to the unique self-limiting growth in the ALD process [21,22]. Because monolayeredness and high conformality are ensured by the process itself, it can be safely assumed that the MoS$_2$ ML is uniform over a one-centimeter scale [23,24]. Hence, the spectroscopic investigation of the electronic structure of MoS$_2$ using a large beam (> 1 mm$^2$ to acquire the interface signals with a reliable statistics) can be realized.

In this study, soft/tender X-ray absorption spectroscopy (XAS) and synchrotron X-ray photoelectron spectroscopy (XPS) were employed to nondestructively inspect the electronic structure of an ALD MoS$_2$ ML film on SiO$_2$/Si. XAS is a technique to measure the X-ray absorption coefficient of a specimen with increasing energy (hν) of incident X-rays. XAS can provide element-specific information on atom's local electronic structure utilizing the strict dipole selection rule for electronic transitions from deep core levels to the unoccupied orbital states. The unoccupied electronic structure for each orbital state can be analyzed by conducting XAS separately at different edges (e.g., Mo $L_3$-edge for Mo $4d$ orbitals, S K-edge for S $3p$ orbitals, etc.). The energies, intensities, and lineshapes of features in the XAS spectra can be interpreted, often being aided by theoretical calculations, to substantiate the details in the hybridized orbital states (such as Mo $4d$–S $3p$ state in $MoS_2$). Particularly, the lowest energy features in the Mo $L_3$-, S K-, and S $L_{2,3}$-edge XAS spectra reflect the electronic structure near the bandgap—that is, the conduction band (CB) information—for Mo $4d$, S $3p$, and S $4s$ orbital states, respectively [25].

On the other hand, XPS is to measure the photoelectron counts at a given hv as a function of electron binding energy (BE). Contrary to XAS, XPS reflects the occupied electronic structure, part of which near the bandgap is the valence band (VB). When a synchrotron XPS is conducted, the VB spectra can be decomposed into the contributions of separate atomic orbitals by utilizing the distinct hv dependencies of the photoionization cross-sections of the atomic orbitals.

The results of the detailed analyses of the XAS and XPS spectra show that the electronic structure of the ALD $MoS_2$ ML is affected by the interfacial S-O interactions at the interface toward $SiO_2$ so that the electronic structure near the bandgap, is constituted not only by the ML itself but also by the hybridized Mo $4d$–S $3p$–O $2p$ states. This strongly suggests that the weak vdW interaction, though not involving the chemical evolution of $Mo^{4+}$ nor $S^{2-}$, can have a substantial impact on determining the electronic interaction of the $MoS_2$ ML.

**Methods**

Monolayer $MoS_2$ in semiconducting 2H phase was prepared via atomic layer deposition on 300 nm-thick $SiO_2$/n-type Si (100) substrates at 175 °C using a four-inch traveling wave-type thermal ALD reactor (CN-1 Co.). $Mo(CO)_6$ and dilute $H_2S$ gas were used as the Mo precursor and sulfur source, respectively. High-purity $N_2$ gas (99.999 %) was used as the purge and carrier gas. The $Mo(CO)_6$ precursor pulse and purge times were 35 and 15 s, respectively. The $H_2S$ pulse and purge times were 3 and 10 s, respectively. The thickness of the $MoS_2$ layer depends on the number of ALD cycles, which was controlled to eight to form the $MoS_2$ ML. Post-deposition annealing (PDA) was performed at 900 °C for 3 min using rapid thermal annealing (RTA, SNTEK Co.) in ambient $H_2S$ for the crystallization of the $MoS_2$ layer. Raman and photoluminescence spectroscopy using a 532-nm laser (customize RM,

Uni-ram) confirmed the formation of a MoS$_2$ ML. The Raman frequency difference between the A$_{1g}$ and E$_{2g}^1$ peaks was 20.1 cm$^{-1}$, which is consistent with literatures [26-28]. The maximum PL intensity was observed at a wavelength of 657.2 nm, corresponding to an energy band gap of ~1.88 eV, consistent with literatures for 2H-MoS$_2$ [26,29,30].

XAS at the Mo L$_3$- and S K-edges (hv ~2.5 keV) was performed at the 16A1 beamline in the Taiwan Light Source in fluorescence yield (FY) mode using a Lytle detector. XAS at the S L$_{2,3}$-edge (hv ~160 eV) was performed at the 4D beamline in the Pohang Light Source (PLS) in partial electron yield (PEY) using a channeltron detector. The sample was rotated by 0º, 45°, and 70° around a vertical axis with respect to the incident X-rays, while the polarization of the X-rays ($E$) was fixed on a horizontal plane. The geometry of the measurements with respect to the sample, is depicted in the upper left corner of Figure 1. The spectrum taken at 0º represents the electronic transition from a core hole to orbital states fully with in-planar orientation, whereas those taken at 45° and 70° undergo the transitions to orbital states with both in-plane and out-of-plane orientations. Since the MoS$_2$ layer is thin (~0.6 nm), the effects of the finite probing depth were negligible. A valence band (VB) XPS analysis with various hv values (90, 200, and 350 eV) was conducted at the 4D beamline in the PLS in the beam normal condition to reveal the orbital characters at the VB.

For the XAS simulation, a full multiple scattering scheme was adopted using an ab-initio real-space multiple scattering code, FEFF (version 8), in which all atoms within an atomic distance of 8 Å from the photon-absorbing atom were considered for the modeled crystal structures (the so-named full multiple scattering (FMS) scheme) [31]. The structure of a free-standing MoS$_2$ ML was taken from Ref. [32], while that of an oxygen-attached ML was generated by adding O atoms on the mirror-reflected sites of Mo in the S-Mo-S structure, thus constructing S-Mo-S-O networks. The dependence of the spectral lineshapes on the detailed position of the added O atom was tested and confirmed to be weak, thereby showing that the S-Mo-S-O model with the specific site can roughly represent most of the cases for the interfacial S-O interactions. The matrix element effects in the initial and final state overlap were fully considered (S$_0^2$ = 1), and the Hedin-Lundqvist self-energy was used for the exchange correlation potential [33]. Next, the angular quantum number ($l$)-projected density of states ($l$DOS) was calculated for comparison with the VB XPS spectra without a core hole [31].

**Results and Discussion**

Figure 1 shows the Mo L$_3$-edge XAS spectra ($2p_{3/2} \rightarrow 4d$) of the MoS$_2$ ML on SiO$_2$ taken at various incidence angles of X-rays ($\theta$ = 0º, 45°, and 70°) with respect to the normal direction of the film. Overall, the lineshapes appear as single peaks at ~2523 eV, which is between the values of the Mo foil (Mo$^0$; ~2521 eV) and the Mo(CO)$_6$ precursor on SiO$_2$ (Mo$^{6+}$; ~2524.5 eV for $t_{2g}$ and ~2527.5 eV for $e_g$), as shown in the appended reference spectra. This is in accordance with the Mo valence of 4+ in the MoS$_2$ ML in that the energy of the first main peak tends to increase as the averaged ionic valence increases [25].

A small but noticeable angle dependence of the main peak was observed; as the $\theta$ (angle between the incident X-ray and the sample normal) increases, the peak shifts to a higher energy (~0.3 eV) and slightly broadens. The theoretical spectra for the *E//ab* and *E//c* polarizations obtained by FEFF calculation for MoS$_2$ ML show very similar behavior to the experimental data in both peak position and broadening, suggesting that the angle dependence indeed reflects the anisotropy in the Mo 4$d$ electronic structure. The simulated spectra for the model including extra O ions (S-Mo-S-O model, which will be discussed later) show polarization dependence almost identical to those for ML, reflecting the nature of locality in the XAS process.

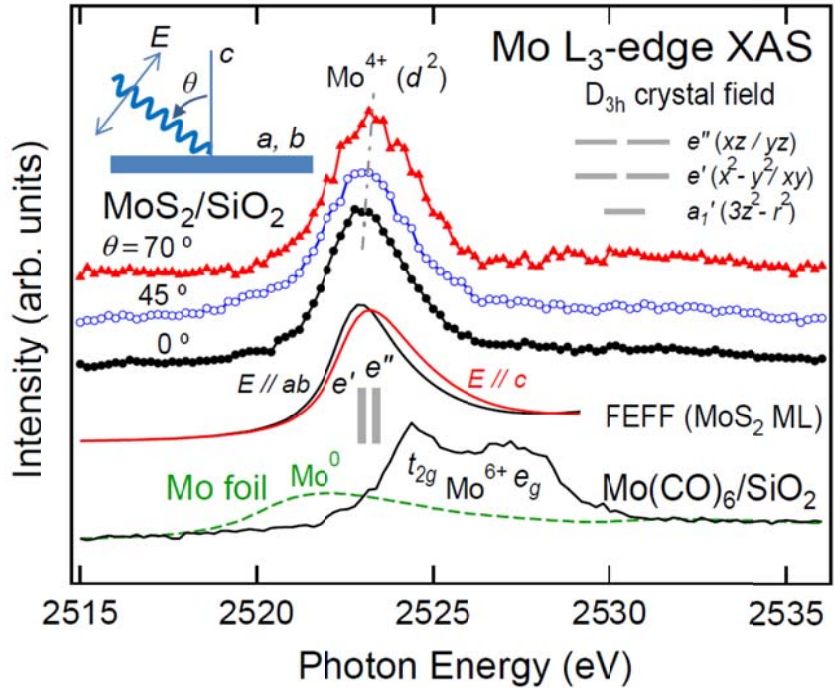

**Figure 1.** Mo $L_3$-edge XAS spectra of $MoS_2$ ML on $SiO_2$ taken at various incidence angles of X-rays ($\theta$ = 0º, 45º, and 70º). Simulated spectra for hypothetical free-standing ML (FEFF) and the spectra of reference Mo foil and $Mo(CO)_6$ precursor were appended for comparison. The angle dependence can be understood by anisotropy in Mo $d^2$ orbital states under $D_{3h}$ local symmetry.

The anisotropic 4$d$ orbital configuration can be understood in terms of ligand field theory. In the $MoS_2$ structure, Mo is surrounded by six S ions with a $D_{3h}$ point group symmetry [34]. Under the symmetry, d orbitals split into $a_1'(3z^2-r^2)$, $e'(x^2-y^2/xy)$, and $e''(xz/yz)$ substates with increasing order of energy. Thus, $d^2$ electrons tend to occupy the $a_1'$ states, preferably with a low spin configuration. However, the strong electron–electron interactions result in a significant mixing of the orbitals so that all the substates are partially filled [10]. Nevertheless, it can be inferred qualitatively that the in-plane $e'(x^2-y^2/xy)$ orbitals have a lower median energy than $e''(xz/yz)$, which implies that the spectrum taken with $E//ab$ would have a lower average peak position than that taken with $E//c$, which is consistent with the experimental data.

As shown in Figure 1, the Mo 4$d$ electronic structure is anisotropic despite the small angle dependence. This is in accordance with the anisotropic nature of the 2-dimensional MoS$_2$ itself [10,35]. Thus, one cannot ascertain at this moment whether the vdW interaction with the SiO$_2$ bottom layer has any influence on the electronic structure of the ML. However, it will be shown in the following figures that the S 3$p$ electronic structure, instead of Mo 4$d$, will show a clear signature of the electronic interaction with O atoms, suggesting that the vdW interaction can induce a substantial modification in the electronic structure, although weak.

Figure 2a shows the S K- (1$s$→ 3$p$) XAS spectra of the MoS$_2$ ML on SiO$_2$. The low-energy region (2469–2474 eV) originates from unoccupied S 3$p$ (due to hybridization with Mo or O orbitals), while the high-energy region (2474–2485 eV) is dominated by a S 3$p$ state that is strongly mixed with S 3$d$ [10]. For comparison, we appended the result of the FEFF simulation on a hypothetical free-standing MoS$_2$ ML, the crystal structure of which was taken from Ref. [32]. In addition, the simulation result for an S-Mo-S-O model in which an O atom is placed below each of the MoS$_2$ units to represent the interfacial vdW interaction is displayed in the figure. A schematic of the configuration is shown in Figure 3.

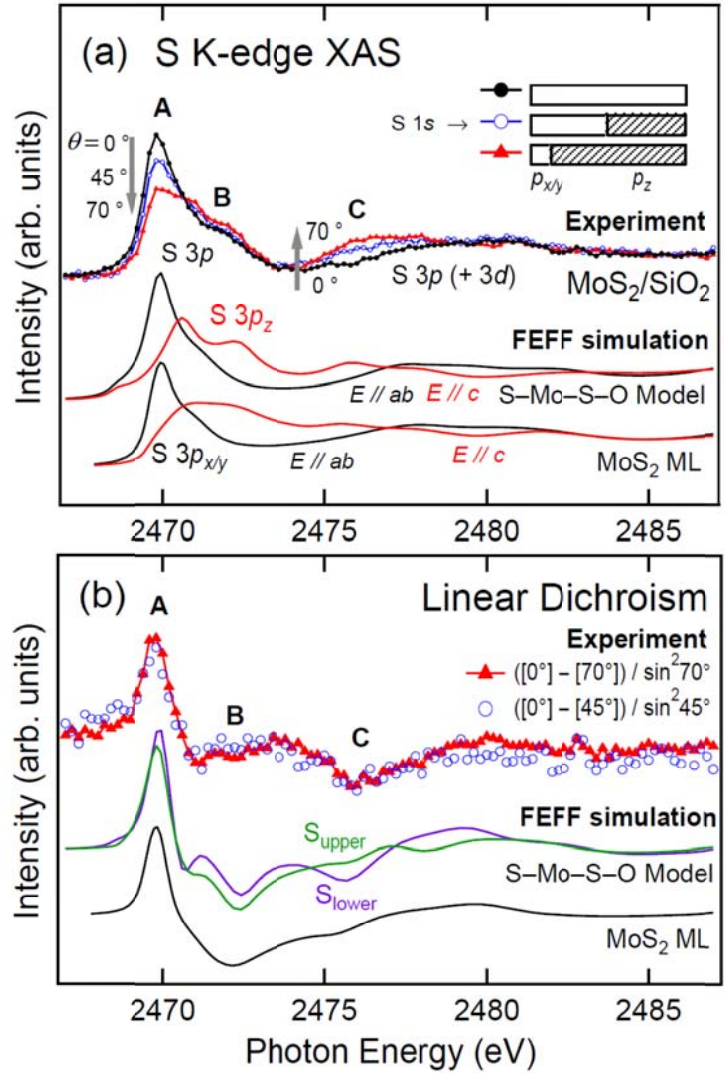

**Figure 2.** (a) S K-edge XAS and (b) LD = $I_{ab} - I_c$ together with results of FEFF simulations on O-appended MoS$_2$ model (S-Mo-S-O) and free-standing ML model. Huge angle dependences are from interfacial S-O interactions as well as the anisotropic electronic structure of MoS$_2$ itself. S$_{upper}$ and S$_{lower}$ stand for S$^{2-}$ ions on top and at interface toward SiO$_2$, respectively. A portion of each $p$ orbital for given $\theta$ is depicted by lengths of bars in inset. See text for more details.

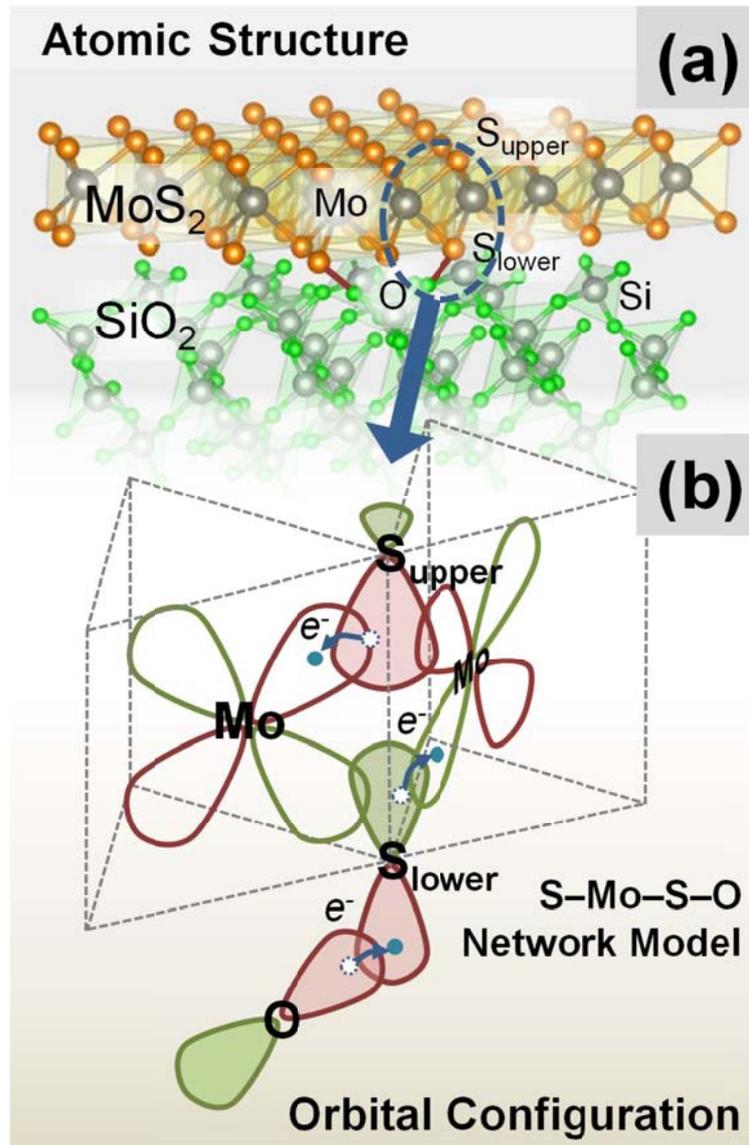

**Figure 3.** (a) Atomic structure of MoS$_2$ ML/SiO$_2$ and (b) orbital configurations according to the S-Mo-S-O model to account for S-O electronic interactions at the MoS$_2$ – SiO$_2$ interface. Charge transfer interactions due to the orbital hybridizations are also depicted.

Overall, the S K-edge XAS lines and their angle dependence appear consistent with the results from both models attached in the figure, while the details are reproduced much more accurately in the case of the S-Mo-S-O model than in the free-standing ML. Three major features were observed: the lowest energy peak at 2470 eV (denoted by A) and higher energy peaks at ~2472 eV (denoted by B) and ~2476 eV (C). The origins of each feature can be identified according to the results of the FEFF simulations. Peaks A and B originate from unoccupied S $3p_{x/y}$ and $3p_z$, respectively, whereas peak C is from S $3p_z$ that is hybridized with S $3d$. These assignments are consistent with the results from a density-functional-theory calculation [10]. Peak A is sharp, reflecting weak hybridization of S $3p_{x/y}$ with Mo $4d$, while peak B is more dispersive than peak A, suggesting a stronger hybridization of S $3p_z$ with Mo $4d$.

It is clearly visible in Figure 2a that the intensities of the features have significant angle dependences; as the $\theta$ increases, the intensity of peak A decreases, while the intensities of peaks B and C increase. Such angle-dependences suggest a strong anisotropy in the hybridized orbital states of the S ions. The $\theta$ dependence of the intensities of peaks A–C can be utilized to scrutinize the anisotropy in the electronic structure of S $3p$ (+ $3d$). The peak intensities ($I$) of the S $1s \rightarrow p$ transition are determined by the unoccupancy of $p$ orbitals ($p_{x/y}$ or $p_z$), abbreviated as [$p_{x/y}$] or [$p_z$], and the orientation of incident X-rays with respect to each of the $p$ orbitals; therefore, $I(\theta)$ = $\cos^2\theta$ [$p_{x/y}$] + $\sin^2\theta$ [$p_z$]. For instance, $\theta = 0°$ data reflects [$p_{x/y}$] only, which corresponds to the theoretical spectra for $E//ab$, while $\theta = 70°$ data reflects 88 % of [$p_z$] and 12 % of [$p_{x/y}$], which correspond mostly to the theoretical spectra for $E//c$. The portions of $p_{x/y}$ and $p_z$ for each $\theta$ are depicted by the lengths of the bars in the legend in Figure 2a.

The strong anisotropy in the S $p$ electronic structure is partly inherited by the crystal structure of the $MoS_2$ ML, as is evidenced by the simulated spectra for the ML (note that the simulation resorts to the crystal structure parameters only). Nevertheless, the trends in the experimental data are not reproduced precisely with $MoS_2$ ML only, thereby evidencing the need for a revised model involving another substance, O, at the $MoS_2/SiO_2$ interface.

It should be noted that peaks B and C are clearly distinguishable by energy, suggesting a clear separation of the quantum states of $3p$ and $3d$. On the other hand, the $p$ and $d$ orbital states under $C_{3v}$ symmetry (no inversion symmetry) are prone to becoming mixed, constituting a broad and mixed orbital state, as shown in the simulated spectra for $MoS_2$ ML at the bottom of Figure 2a. Therefore, the separation of peaks B and C indicates that the mixing of the S $p$ and $d$ states has become weak, which is plausible when the $S^{2-}$ ions are positioned in a different coordination environment augmented by adjacent $Si^{4+}$ or $O^{2-}$ ions in the bottom $SiO_2$ layer. Figure 2a clearly shows that the separation is well reproduced in the simulated spectra for the S-Mo-S-O model. Indeed, the broadening and small separation of peaks B and C were observed experimentally in the XAS spectra of an ultrathin $MoS_2$ [36].

To investigate the origin of the orbital anisotropy more clearly, the linear dichroism (LD) —which is the difference between the spectral intensities for $E//ab$ ($I_{ab}$) and $E//c$ ($I_c$), that is, LD = $I_{ab}$ - $I_c$— of the $MoS_2/SiO_2$ sample is displayed in Figure 2b. The LD obtained by subtracting the $\theta = 70°$ data from the $\theta = 0°$ data is very similar to that obtained by subtracting the $\theta = 45°$ data from the $\theta = 0°$ data, clearly showing that $\theta$ dependence is systematic.

The lineshapes of dips B and C are of particular interest. The simulated LD for the hypothetical $MoS_2$ ML does not show well-separated dip features as it does in the experimental data, whereas that for the S-Mo-S-O model does. In the S-Mo-S-O model, the S ions facing the bottom $SiO_2$ (denoted by $S_{lower}$) have distinct local coordination (with additional O ions at the bottom; see Figure 3b) from the S ions on the top (denoted by $S_{upper}$). Thus, the simulated LDs of $S_{lower}$ and $S_{upper}$ are different, as shown in Figure 2b. The details of the dip features in the LD of $S_{lower}$ are highly consistent with those in the experimental data. This strongly suggests that the inclusion of O in S coordination is essential for understanding the electronic structure of the $MoS_2/SiO_2$ interface. Meanwhile, those in the LD of $S_{upper}$ appear as a mixture of LDs of $S_{lower}$ and of S ions in the free-standing ML, which is reasonable in the sense that the impact of the interfacial interaction would not perfectly reach the top S ions.

Dip C in the LD of $S_{lower}$ is particularly prominent, implying that the $p$-$d$ mixing becomes weak owing to the breaking of the $C_{3v}$ symmetry, thereby reinstating the separated S $p$ and $d$ orbital states. Thus, peak C in Figure 2a and dip C in Figure 2b are indicative of the interfacial reaction. Because the displacement of the adjacent O ions with respect to $S_{lower}$ would be mostly along the direction of the c-axis, the signature of the interfacial reaction should be more prominent in the $p_z$ signal ($E//c$) than in the $p_{x/y}$ signal ($E//ab$), which are hybridized (though weakly) with S $3d$.

Figure 3 summarizes the results of the analyses of the S K-edge XAS and LD schematically. Figure 3a schematically shows the atomic structures of $MoS_2$ $ML/SiO_2$, the atomic coordinates of which are taken from a bulk $MoS_2$ [37] and an α-quartz $SiO_2$ [38]. The substantial contribution of weak S-O bonds to the electronic interaction is depicted as thick bars between the S (in $MoS_2$) and O (in $SiO_2$) atoms. The orbital configurations near a given S atom are displayed as lobes in Figure 3b (red and green are for the + and - signs of the orbital wavefunctions, respectively). Two Mo 4$d$ (empty) orbitals out of the three near the S atom are displayed for convenience, together with the S$_{upper}$ 3$p$, S$_{lower}$ 3$p$, and O 2$p$ orbitals (filled orbitals). As shown in Fig. 2b, S 3$p_z$ (rather than S 3$p_{x/y}$) is strongly hybridized with Mo 4$d$ owing to the inherent 2D nature of MoS$_2$ with preferential hybridization along the z-direction. Thus the S orbitals are displayed as vertical lobes representatively. In the case of S$_{upper}$, the intra-atomic $p$-$d$ mixing is strong, resulting in asymmetrical lobes. Meanwhile, in the case of S$_{lower}$, the inclusion of O 2$p$ – S$_{lower}$ 3$p_z$ hybridization weakens the $p$-$d$ mixing to reinstate the $p$ symmetry (see Fig. 2b).

The charge transfer (CT) from higher electronegativity ions to lower electronegativity ions (S$_{upper}$/S$_{lower}$ → Mo and O → S$_{upper}$) is depicted as well. Due to the anisotropic orbital configurations, CTs prefer to occur along the z-direction along with the network of S$_{upper}$–Mo–S$_{lower}$(–O). In the case of S$_{upper}$, an in-plane CT though Mo $4d$ – S$_{upper}$ 3$p_z$ hybridization channels in the 2D structure will be dominated. On the other hand, in the case of S$_{lower}$, the inclusion of O 2$p$ in the CT network will contribute the electrical conduction via the hybridized O 2$p$ – S$_{lower}$ 3$p_z$ states at the interface. Therefore, the electronic structure of MoS$_2$ ML can be modified by the presence of SiO$_2$ due to the interfacial vdW interaction.

The influences of the interfacial vdW interaction on the unoccupied and occupied electronic structures are demonstrated in Figures 4 and 5. Figure 4 shows the S L$_{2,3}$-edge (2$p$ → 3$d$/4$s$) XAS spectra, in which the contributions from the 2$p_{3/2}$ and 2$p_{1/2}$ initial states overlap with a separation in energy of ~1.2 eV [10]. Compared to S K-edge XAS, S L$_{2,3}$-edge XAS has an advantage in investigating the electronic structure with a minimal photoexcitation effect because the S 2$p$ core level is relatively shallow (~160 eV) and has a higher angular momentum than S 1$s$ (~2470 eV); thus, the XAS process would have a stronger core-hole screening effect [31]. As a result, the S L$_{2,3}$-edge XAS spectra could be compared directly to the unoccupied electronic structure in the ground state, except for the duplication of L$_3$- and L$_2$-edge features.

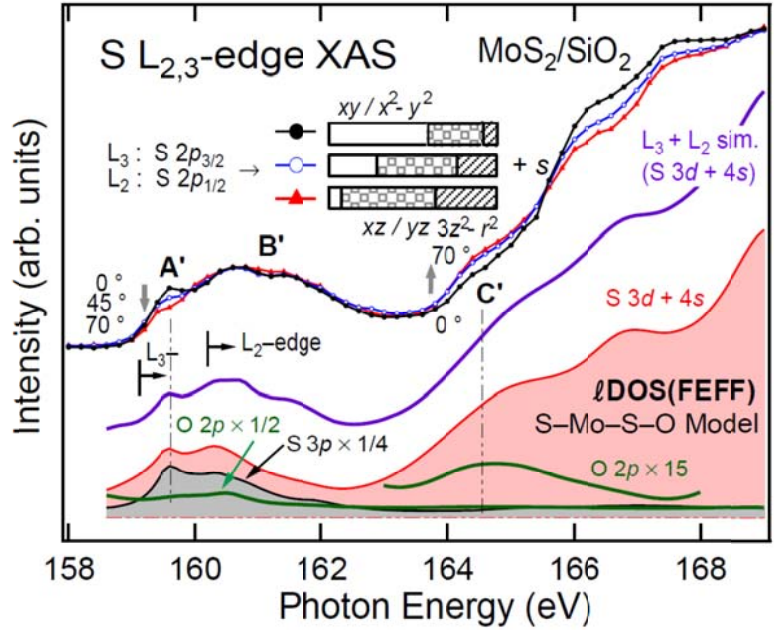

**Figure 4.** S L$_{2,3}$-edge XAS spectra of MoS$_2$ ML/SiO$_2$ along with calculated ℓDOSs (FEFF) and simulated spectrum for the S-Mo-S-O model. Portion of each $d$ orbital for given $\theta$ is depicted by lengths of bars in inset. The contribution of O $p$ DOS to CB as well as peak C is visible, suggesting a substantial influence of O inclusion on unoccupied electronic structure.

For comparison, the calculated density of states (DOS) projected to the specific angular quantum number (ℓDOS) from FEFF for the S-Mo-S-O model is included in the figure. Here, the energy of the ℓDOSs was shifted arbitrarily to compensate for the S $2p_{3/2}$ core level energy. Based on the calculated S $d$ and $s$ DOSs, a theoretical spectrum was constructed (assuming complete core hole screening) as shown in the figure as well [31]. Because the simulated spectrum reproduces the overall lineshapes in the experimental spectra very well, the features in the experimental spectra could be assigned according to the ℓDOSs.

It was observed that the DOS in the low-energy region (159–163 eV) dominated by S $3d$ mixed with $4s$, appears similar to the S $p$ DOS (reduced to 1/4). The similarity reflects the effect of $s$-$p$-$d$ orbital mixing in the S ions under C$_{3v}$ symmetry. Owing to the intra-atomic orbital mixing, the angle dependence in the S K-edge XAS spectra ($3p$) reappeared in the S L-edge spectra. The features denoted by A' to C' in Figure 4 correspond to A to C in Figure 2, respectively. The major variations in the intensities of A' to C' with increasing $\theta$ are consistent with those of A to C.

The portions of the $2p \rightarrow 3d$ transition matrix elements ($L_{2,3}$-edge) for each $\theta$ are denoted by the lengths of the bars in the legend. The $\theta$ dependences of the $d$ orbitals are different from those of the $p$ orbitals (K-edge) because the $d$ shell comprises ($x^2$-$y^2$/$xy$), ($xz$/$yz$), and ($3z^2$-$r^2$) of markedly different symmetry from $p_{x/y}$ or $p_z$. Nevertheless, it can be said that they are similar to each other in that the signals from the orbitals along the $c$-axis become more intense as $\theta$ increases. Therefore, the weakening of peak A' and the enhancement of peak C' with increasing $\theta$, can be understood by the same origins as the case of peaks A and C in Figure 2a, which is both the anisotropy of the interfacial interaction and that of the MoS$_2$ ML itself.

It should be noted that the unoccupied O $p$ DOS, which cannot be directly observed in S K- or S L-edge XAS, also have similar peak features, as highlighted by the vertical lines in Figure 4. The similarity is a consequence of the involvement of weak S–O electronic interactions. Particularly for peak C', the O $p$ DOS at ~164 eV protrudes, as is highlighted in the magnified plot in the right bottom panel, implying that the interfacial interaction indeed contributes to peak C' (as well as peak C). Furthermore, the O $p$ DOS has a non-negligible contribution in the low-energy region, suggesting that it can interfere with the CB of MoS$_2$ to contribute to electrical conduction.

Figure 5 shows the XPS VBs taken at three different photon energies: hν = 90 eV, 200 eV, and 350 eV. The hν-dependent XPS is often employed to scrutinize the depth profile utilizing the different inelastic mean free paths of the photoelectrons [39]. However, the probing depth issue is irrelevant to the case of MoS$_2$ ML/SiO$_2$ because the thickness of the ML is far less than the general range of the probing depth (~2 nm) [40]. In contrast, the sensitivity of each orbital state changes greatly depending on the respective photoionization cross-section [41]. Particularly for Mo 4$d$ orbitals, there exists a Cooper minimum of approximately hν = 90 eV, so the Mo 4$d$ DOS is invisible in the spectrum measured at the photon energy [42,43].

The VB of MoS$_2$ ML consists of Mo 4$d$ (near the Fermi level, E$_F$), S 3$p$ (BE< 10 eV), and S 3$s$ (BE ~14 eV) [26,43]. Because Mo$^{4+}$ ions possess a $d^2$ configuration (see Figure 1), the DOS near the VB maximum (denoted by peaks α and β) should be dominated by Mo 4$d$ states constituting a direct Mott gap (~1.8 eV) [44-46], which is in accordance with the photoluminescence energy. It follows that the contributions of S$^{2-}$ ($p^6$) appear at higher BEs (peaks γ and δ). In addition, owing to the presence of SiO$_2$ at the bottom of MoS$_2$, the signatures of Si 3$p$ + O 2$p$ in the region of a very large BE (5–16 eV) would appear as in the appended SiO$_2$ reference spectrum [47]. If there is no interaction between MoS$_2$ and SiO$_2$, SiO$_2$ will remain highly insulating, and therefore, there is no chance for the O 2$p$ DOS to exist near the E$_F$.

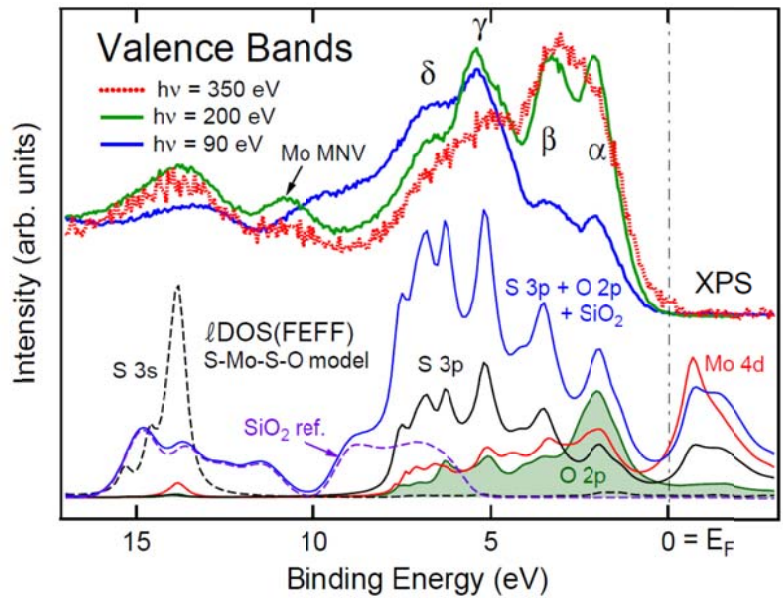

**Figure 5.** VB XPS spectra taken at various photon energies to scrutinize orbital characters. ℓ-projected DOSs for S-Mo-S-O model together with SiO$_2$ reference spectrum are appended. Evolution in intensity of peaks α-δ with increasing photon energy reflects O 2$p$ DOS at VB, suggesting substantial contribution of S-O interfacial interactions on the occupied electronic structure.

All the peaks in the experimental spectra (denoted by peaks α to δ in particular) can be assigned according to the ℓDOSs from the S-Mo-S-O model (appended in the figure), except for the Mo MNV Auger line in the hν = 200 eV data. For the hν = 90 eV data, only the contributions from SiO$_2$, S 3$p$, and O 2$p$ appear because Mo 4$d$ is invisible, which results in reduced intensity of peaks α and β. As hν increases (200 eV and 350 eV), on the other hand, the Mo 4$d$ intensity becomes stronger so that peaks α and β protrude compared to the hν = 90 eV data.

Interestingly, it was observed in the hν = 350 eV data that peak α becomes weaker than in the hν = 200 eV data, whereas peak β does not. If peak α originated purely from Mo 4*d*, it could not be weakened significantly while leaving peak β still intense. A lower energy resolution of the hν = 350 eV data cannot account for the selective weakening of peak α. This implies that peak α is contributed to by some orbital states other than from $MoS_2$ itself.

The O *p* DOS from the S-Mo-S-O model exists near the $E_F$ with its peak features similar to the S 3*p* DOS due to S 3*p*–O 2*p* hybridization (see Figure 3). The O *p* DOS has maximal intensity at the energy of peak α. Therefore, the contribution of O 2*p* to peak α can allow a distinct hν-dependence of peak α compared to peak β. Indeed, the ratio of the photoionization cross-section between O 2*p* and Mo 4*d* drops rapidly from 0.72 (at 200 eV) to 0.25 (at 350 eV) [41], so that the contribution of O 2p in peak β will appear much weaker in the hν = 350 eV data.

Therefore, it can be concluded that the experimental VBs of $MoS_2/SiO_2$ have a fingerprint of O 2*p* contribution at the VB maximum. Together with the findings shown in Figure 4, the contribution of O 2*p* in the CB, this finding suggests the possibility of parasitic conduction through the O 2*p* near $S_{lower}$ at the interface of $MoS_2$ and $SiO_2$. This is in contrast to the common belief that the influence of the vdW interaction between the semiconducting $MoS_2$ and highly insulating $SiO_2$ on the electronic structure is negligible [19,48]. However, there have been reports that certain charge trapping can occur in $SiO_2$ layers in the vicinity of $MoS_2$ [49,50], which might imply that electron hopping among adjacent S-O ions is feasible.

The $MoS_2$ ML/$SiO_2$ deposited by ALD, which is known for utmost stability in the growth of binary oxides, exhibits a slight but substantial electronic distinction from the ideal free-standing $MoS_2$ owing to interfacial S–O interactions. It is noteworthy that the $Mo^{4+}$ and $S^{2-}$ ions nevertheless remain mostly intact in that the peak positions and lineshapes of the first peaks (e'/e" in Figure 1, A in Figure 2 and A' in Figure 4) coincide with the theoretical expectations for a $MoS_2$ ML. Parija *et al.* reported that certain pre-edge states at energies lower than the main peak, which is particularly prominent at the edges of the $MoS_2$ nanostructures, are responsible for the catalytic activity of $MoS_2$ [3,10,11]. In ALD $MoS_2$, however, the volume of such edges is minimal compared to the main body because of the excellent uniformity properties of ALD films over large areas. Therefore, the difference in the chemistry of the ALD $MoS_2$ ML/$SiO_2$ from ideal free-standing $MoS_2$ is not significant. Hence, the interfacial interaction is electronic.

**Conclusion**

The influence of the electronic interaction at the $MoS_2$–$SiO_2$ interface was substantiated by a combination of X-ray spectroscopy (XAS and XPS) and ab initio calculation (FEFF) for an ALD $MoS_2$ ML film on $SiO_2$. Detailed spectral analyses of the angle-dependent XAS and photon-energy-dependent XPS data, along with the comparison with the calculation results, clearly show that the $O^{2-}$ ions in $SiO_2$ adjacent to $S^{2-}$ in $MoS_2$ ML can participate in the orbital hybridization so that O 2$p$ can contribute to both the valence and conduction bands of the system, despite the highly insulating nature of $SiO_2$. This finding is direct evidence of the substantial electronic influence of the vdW coupling at the $MoS_2/SiO_2$ interface, which is essential for a precise understanding of the electronic structure of $MoS_2$-based devices.

AUTHOR INFORMATION

**Corresponding Author**

*The author to whom correspondence should be addressed: zax@jbnu.ac.kr (D.-Y. Cho)

**Author Contributions**

The manuscript was written through contributions of all authors. All authors have given approval to the final version of the manuscript.

ACKNOWLEDGMENT

This work was supported by Basic Science Research Program (Grant No. 2018R1D1A1B07043427) through the National Research Foundation of Korea funded by the Korea government Ministry of Education, and the Future Semiconductor Device Technology Development Program (20010558) funded by Ministry of Trade, Industry & Energy (MOTIE) and Korea Semiconductor Research Consortium (KSRC).

REFERENCES

(1) Hinnemann, B.; Moses, P. G.; Bonde, J.; Jørgensen, K. P.; Nielsen, J. H.; Horch, S.; Chorkendorff, I.; Nørskov, J. K. Biomimetic Hydrogen Evolution: $MoS_2$ Nanoparticles as Catalyst for Hydrogen Evolution. *J. Am. Chem. Soc.* **2005,** *127*, 5308-5309.

(2) Voiry, D.; Salehi, M.; Silva, R.; Fujita, T.; Chen, M.; Asefa, T.; Shenoy, V. B.; Eda, G.; Chhowalla, M. Conducting $MoS_2$ Nanosheets as Catalysts for Hydrogen Evolution Reaction. *Nano Lett.* **2013,** *13*, 6222-6227.

(3) Jaramillo, T. F.; Jørgensen, K. P.; Bonde, J.; Nielsen, J. H.; Horch, S.; Chorkendorff, I. Identification of Active Edge Sites for Electrochemical $H_2$ Evolution from $MoS_2$ Nanocatalysts. *Science* **2007,** *317*, 100-102.

(4) Ding, Q.; Meng, F.; English, C. R.; Cabán-Acevedo, M.; Shearer, M. J.; Liang, D.; Daniel, A. S.; Hamers, R. J.; Jin, S. Efficient Photoelectrochemical Hydrogen Generation using Heterostructures of Si and Chemically Exfoliated Metallic $MoS_2$. *J. Am. Chem. Soc.* **2014,** *136*, 8504-8507.

(5) Kwon, K. C.; Choi, S.; Hong, K.; Moon, C. W.; Shim, Y. S.; Kim, T.; Sohn, W.; Jeon, J. M.; Lee, C.-H.; Nam, K. T.; Han, S.; Kim, S. Y.; Jang, H. W. Wafer-Scale Transferable Molybdenum Disulfide Thin-Film Catalysts for Photoelectrochemical Hydrogen Production. *Energy Environ. Sci.* **2016,** *9*, 2240-2248.

(6) Merki, D.; Fierro, S.; Vrubel, H.; Hu, X. Amorphous Molybdenum Sulfide Films as Catalysts for Electrochemical Hydrogen Production in Water. *Chem. Sci.* **2011,** *2*, 1262-1267.

(7) Lu, A. Y.; Yang, X.; Tseng, C. C.; Min, S.; Lin, S. H.; Hsu, C. L.; Li, H.; Idriss, H.; Kuo, J. L.; Huang, K. W. Li, L. J. High-Sulfur-Vacancy Amorphous Molybdenum Sulfide as a High Current Electrocatalyst in Hydrogen Evolution. *Small* **2016,** *12*, 5530-5537.

(8) Vrubel, H.; Hu, X. Growth and Activation of an Amorphous Molybdenum Sulfide Hydrogen Evolving Catalyst. *ACS Catal.* **2013,** *3*, 2002-2011.

(9) Zhou, J.; Dai, S.; Dong, W.; Su, X.; Fang, L.; Zheng, F.; Wang, X.; Shen, M. Efficient and Stable $MoS_2$ Catalyst Integrated on Si Photocathodes by Photoreduction and Post-Annealing for Water Splitting. *Appl. Phys. Lett.* **2016,** *108*, 213905.

(10) Parija, A.; Choi, Y.-H.; Liu, Z.; Andrews, J. L.; De Jesus, L. R.; Fakra, S. C.; Al-Hashimi, M.; Batteas, J. D.; Prendergast, D.; Banerjee, S. Mapping Catalytically Relevant Edge Electronic States of $MoS_2$. *ACS Cent. Sci.* **2018,** *4*, 493-503.

(11) Zhou, W.; Zou, X.; Najmaei, S.; Liu, Z.; Shi, Y.; Kong, J.; Lou, J.; Ajayan, P. M.; Yakobson, B. I.; Idrobo, J. C. Intrinsic Structural Defects in Monolayer Molybdenum Disulfide. *Nano Lett.* **2013,** *13*, 2615-2622.

(12) Lukowski, M. A.; Daniel, A. S.; Meng, F.; Forticaux, A.; Li, L.; Jin, S. Enhanced Hydrogen Evolution Catalysis from Chemically Exfoliated Metallic $MoS_2$ Nanosheets. *J. Am. Chem. Soc.* **2013,** *135*, 10274-10277.

(13) Tang, Q.; Jiang, D. E. Mechanism of Hydrogen Evolution Reaction on 1T-$MoS_2$ from First Principles. *ACS Catal.* **2016,** *6*, 4953-4961.

(14) Liu, Q.; Li, X.; He, Q.; Khalil, A.; Liu, D.; Xiang, T.; Wu, X.; Song, L. Gram-scale Aqueous Synthesis of Stable Few-layered 1T-MoS$_2$: Applications for Visible-Light-Driven Photocatalytic Hydrogen Evolution. *Small* **2015,** *11*, 5556-5564.

(15) Zhang, W.; Liao, X.; Pan, X.; Yan, M.; Li, Y.; Tian, X.; Zhao, Y.; Xu, L.; Mai, L. Superior Hydrogen Evolution Reaction Performance in 2H-MoS$_2$ to that of 1T Phase. *Small* **2019,** *15*, 1900964.

(16) Maitra, U.; Gupta, U.; De, M.; Datta, R.; Govindaraj, A.; Rao, C. N. R. Highly Effective Visible-Light-Induced H$_2$ Generation by Single-layer 1T-MoS$_2$ and a Nanocomposite of Few-Layer 2H-MoS$_2$ with Heavily Nitrogenated Graphene. *Angew. Chem. Int. Ed.* **2013,** *52*, 13057-13061.

(17) Hao, L.; Gao, W.; Liu, Y.; Han, Z.; Xue, Q.; Guo, W.; Zhu, J.; Li, Y. R. High-Performance n-MoS$_2$/i-SiO$_2$/p-Si Heterojunction Solar Cells. *Nanoscale* **2015,** *7*, 8304-8308.

(18) Hao, L.; Liu, Y.; Han, Z.; Xu, Z.; Zhu, J. Giant Lateral Photovoltaic Effect in MoS$_2$/SiO$_2$/Si Pin Junction. *J. Alloys Compd.* **2018,** *735*, 88-97.

(19) Cheng, Y.; Yao, K.; Yang, Y.; Li, L.; Yao, Y.; Wang, Q.; Zhang, X.; Han, Y.; Schwingenschlögl, U. Van der Waals Epitaxial Growth of MoS$_2$ on SiO$_2$/Si by Chemical Vapor Deposition. *RSC Adv.* **2013,** *3*, 17287-17293.

(20) Lee, Y. H.; Zhang, X. Q.; Zhang, W.; Chang, M. T.; Lin, C. T.; Chang, K. D.; Yu, Y. C.; Wang, J. T. W.; Chang, C. S.; Li, L. J.; Lin T. W. Synthesis of Large-Area MoS$_2$ Atomic Layers with Chemical Vapor Deposition. *Adv. Mater.* **2012,** *24*, 2320-2325.

(21) Johnson, R. W.; Hultqvist, A.; Bent, S. F. A Brief Review of Atomic Layer Deposition: from Fundamentals to Applications. *Mater. Today* **2014,** *17*, 236-246.

(22) George, S. M. Atomic Layer Deposition: an Overview. *Chem. Rev.* **2010,** *110*, 111-131.

(23) Pyeon, J. J.; Kim, S. H.; Jeong, D. S.; Baek, S.-H.; Kang, C.Y.; Kim, J.-S.; Kim, S. K. Wafer-Scale Growth of MoS$_2$ Thin Films by Atomic Layer Deposition. *Nanoscale* **2016,** *8*, 10792-10798.

(24) Letourneau, S.; Young, M. J.; Bedford, N. M.; Ren, Y.; Yanguas-Gil, A.; Mane, A. U.; Elam, J. W.; Graugnard, E. Structural Evolution of Molybdenum Disulfide Prepared by Atomic Layer Deposition for Realization of Large Scale Films in Microelectronic Applications. *ACS Appl. Nano Mater.* **2018,** *1*, 4028-4037.

(25) De Groot, F.; Kotani, A. *Core Level Spectroscopy of Solids*. CRC press: Boca Raton, Florida, 2008.

(26) Tao, J.; Chai, J.; Lu, X.; Wong, L. M.; Wong, T. I.; Pan, J.; Xiong, Q.; Chi, D.; Wang, S. Growth of Wafer-Scale $MoS_2$ Monolayer by Magnetron Sputtering. *Nanoscale* **2015,** *7*, 2497-2503.

(27) Bagnall, A.; Liang, W.; Marseglia, E.; Welber, B. Raman Studies of $MoS_2$ at High Pressure. *Physica B+C* **1980,** *99*, 343-346.

(28) Wu, J. B.; Zhao, H.; Li, Y.; Ohlberg, D.; Shi, W.; Wu, W.; Wang, H.; Tan, P. H. Monolayer Molybdenum Disulfide Nanoribbons with High Optical Anisotropy. *Adv. Opt. Mater.* **2016,** *4*, 756-762.

(29) Nayak, A. P.; Pandey, T.; Voiry, D.; Liu, J.; Moran, S. T.; Sharma, A.; Tan, C.; Chen, C. H.; Li, L.-J.; Chhowalla, M. Pressure-Dependent Optical and Vibrational Properties of Monolayer Molybdenum Disulfide. *Nano Lett.* **2015,** *15*, 346-353.

(30) Song, I.; Park, C.; Choi, H. C. Synthesis and Properties of Molybdenum Disulphide: from Bulk to Atomic Layers. *RSC Adv.* **2015,** *5*, 7495-7514.

(31) Ankudinov, A. L.; Ravel, B.; Rehr, J. J.; Conradson, S. D. Real-Space Multiple-Scattering Calculation and Interpretation of X-Ray-Absorption Near-Edge Structure. *Phys. Rev. B* **1998,** *58*, 7565-7576.

(32) Blumberg, A.; Keshet, U.; Zaltsman, I.; Hod, O. Interlayer Registry to Determine the Sliding Potential of Layered Metal Dichalcogenides: the case of 2H-$MoS_2$. *J. Phys. Chem. Lett.* **2012,** *3*, 1936-1940.

(33) Rehr, J. J.; Albers, R. C. Theoretical Approaches to X-Ray Absorption Fine Structure. *Rev. Mod. Phys.* **2000,** *72*, 621-654.

(34) Lince, J. R.; Didziulis, S. V.; Yarmoff, J. A. Resonant Photoelectron Spectroscopy at the Mo 4p→ 4d Absorption Edge in $MoS_2$. *Phys. Rev. B* **1991,** *43*, 4641-4647.

(35) Li, D.; Bancroft, G.; Kasrai, M.; Fleet, M.; Feng, X.; Tan, K. Polarized X-Ray Absorption Spectra and Electronic Structure of Molybdenite (2H-MoS$_2$). *Phys. Chem. Miner.* **1995**, *22*, 123-128.

(36) Guay, D.; Divigalpitiya, W.; Belanger, D.; Feng, X. Chemical Bonding in Restacked Single-Layer MoS$_2$ by X-Ray Absorption Spectroscopy. *Chem. Mater.* **1994**, *6*, 614-619.

(37) Anghel, S.; Chumakov, Y.; Kravtsov, V.; Volodina, G.; Mitioglu, A.; Płochocka, P.; Sushkevich, K.; Mishina, E.; Kulyuk, L. Site-Selective Luminescence Spectroscopy of Bound Excitons and Local Band Structure of Chlorine Intercalated 2H-and 3R-MoS$_2$ Polytypes. *J. Lumin.* **2016,** *177*, 331-336.

(38) Di Pomponio, A.; Continenza, A. Structural Properties of α-Quartz under High Pressure and Amorphization Effects. *Phys. Rev. B* **1993**, *48*, 12558-12565.

(39) Woicik, J. C.; Woicik, J. *Hard X-Ray Photoelectron Spectroscopy (HAXPES)*. Springer: Berlin, **2015**.

(40) Hüfner, S., *Photoelectron Spectroscopy: Principles and Applications*. Springer: Berlin, **2003**.

(41) Yeh, J.; Lindau, I. Atomic Subshell Photoionization Cross Sections and Asymmetry Parameters: 1⩽ Z⩽ 103. *At. Data Nucl. Data Tables* **1985**, *32*, 1-155.

(42) Fleischauer, P. D.; Lince, J. R.; Bertrand, P.; Bauer, R. Electronic Structure and Lubrication Properties of Molybdenum Disulfide: A Qualitative Molecular Orbital Approach. *Langmuir* **1989,** *5*, 1009-1015.

(43) Han, S. W.; Cha, G. B.; Frantzeskakis, E.; Razado-Colambo, I.; Avila, J.; Park, Y. S.; Kim, D.; Hwang, J.; Kang, J. S.; Ryu, S. Band-Gap Expansion in the Surface-Localized Electronic Structure of MoS$_2$ (0002). *Phys. Rev. B* **2012,** *86*, 115105.

(44) Tang, Q.; Jiang, D. E. Stabilization and Band-Gap Tuning of the 1T-MoS$_2$ Monolayer by Covalent Functionalization. *Chem. Mater.* **2015,** *27*, 3743-3748.

(45) Radisavljevic, B.; Radenovic, A.; Brivio, J.; Giacometti, V.; Kis, A. Single-Layer MoS$_2$ Transistors. *Nat. Nanotechnol.* **2011,** *6*, 147-150.

(46) Eknapakul, T.; King, P.; Asakawa, M.; Buaphet, P.; He, R.-H.; Mo, S.-K.; Takagi, H.; Shen, K.; Baumberger, F.; Sasagawa, T. Electronic Structure of a Quasi-Freestanding MoS$_2$ Monolayer. *Nano Lett.* **2014,** *14*, 1312-1316.

(47) Zakaznova-Herzog, V. P.; Nesbitt, H.; Bancroft, G.; Tse, J.; Gao, X.; Skinner, W. High-Resolution Valence-Band XPS Spectra of the Nonconductors Quartz and Olivine. *Phys. Rev. B* **2005,** *72*, 205113.

(48) Su, X.; Cui, H.; Ju, W.; Yong, Y.; Li, X. *Mod.* First-Principles Investigation of MoS$_2$ Monolayer Adsorbed on SiO$_2$ (0001) Surface. *Phys. Lett. B* **2017,** *31*, 1750229.

(49) Shlyakhov, I.; Chai, J.; Yang, M.; Wang, S.; Afanas' ev, V.; Houssa, M.; Stesmans, A. Band Alignment at Interfaces of Synthetic Few-Monolayer MoS$_2$ with SiO$_2$ from Internal Photoemission. *APL Mater.* **2018,** *6*, 026801.

(50) Guo, Y.; Wei, X.; Shu, J.; Liu, B.; Yin, J.; Guan, C.; Han, Y.; Gao, S.; Chen, Q. Charge Trapping at the MoS$_2$-SiO$_2$ Interface and its Effects on the Characteristics of MoS$_2$ Metal-Oxide-Semiconductor Field Effect Transistors. *Appl. Phys. Lett.* **2015,** *106*, 103109.